\documentclass[american,preprint]{aastex6}
\usepackage[T1]{fontenc}
\usepackage[latin9]{inputenc}
\usepackage{graphicx}

\slugcomment{Submitted to The Astronomical Journal}
\shorttitle{Masses of Kepler-46b,c from TTVs}
\shortauthors{Saad-Olivera et al.}

\usepackage{babel}

\begin{document}

\title{Masses of Kepler-46b,c from Transit Timing Variations}

\author{Ximena Saad-Olivera\altaffilmark{1}, David Nesvorn\'{y}\altaffilmark{1,2}, David Kipping\altaffilmark{3}, and Fernando Roig\altaffilmark{1}} 
\affil{\altaffilmark{1}{Observat\'{o}rio Nacional, Rio de Janeiro, Jos\'e Cristino 77, Rio de Janeiro, RJ 20921-400, Brazil}\\
\altaffilmark{2}{Department of Space Studies, Southwest Research Institute, 1050 Walnut Street,Suite 300, Boulder, CO 80302, USA}\\
\altaffilmark{3}{Dept. of Astronomy, Columbia University, 550 W 120th St., New York, NY 10027, USA}}

\altaffiltext{1}{ximena@on.br}

\begin{abstract}

We use 16 quarters of the \textit{Kepler} mission data to analyze the transit timing 
variations (TTVs) of the extrasolar planet Kepler-46b (KOI-872). Our dynamical 
fits confirm that the TTVs of this planet (period $P=33.648^{+0.004}_{-0.005}$ days) 
are produced by a non-transiting planet Kepler-46c ($P=57.325^{+0.116}_{-0.098}$ days). 
The Bayesian inference tool \texttt{MultiNest} is used to infer the dynamical 
parameters of Kepler-46b and Kepler-46c. We find that the two planets have nearly 
coplanar and circular orbits, with eccentricities $\simeq 0.03$ somewhat higher than 
previously estimated. The masses of the two planets are found to be 
$M_{b}=0.885^{+0.374}_{-0.343}$ and 
$M_{c}=0.362^{+0.016}_{-0.016}$ Jupiter masses, 
with $M_{b}$ being determined here from TTVs for the first time. Due to the 
precession of its orbital plane, Kepler-46c should start transiting its 
host star in a few decades from now.

\end{abstract}


\section{Introduction} \label{sec:intro}
More than 3300 exoplanets have been discovered with the \textit{Kepler} and \textit{K2} 
missions, and almost 16\% of these are in multi-transiting planetary 
systems with at least two confirmed planets
(\url{exoplanetarchive.ipac.caltech.edu})
The detection of these exoplanets was possible because they have nearly edge-on orbits 
and pass in front of the star, thus producing a dip in the photometric 
light curve. The analysis of the transit light curve allows scientists 
to determine the planetary to stellar radius ratio (${R_P}/{R_{\star}}$), 
the inclination of the planetary orbit ($i_P$), the transit duration ($t_D$), 
and the time of mid-transit ($t_C$) (e.g., Seager \& Mall\'{e}n-Ornelas 2003, 
Carter et al. 2008). 

In multi-planetary systems, the mutual gravitational perturbation between 
planets can be detectable under certain conditions. For example, the mid-transit 
time $t_C$ of consecutive transits of the same planet may not occur 
exactly on constant ephemeris. Instead, due to the gravitational perturbations 
from planetary companions, $t_C$ values can display non linear trends called 
Transit Timing Variations (TTVs). In some cases, it is also possible to measure 
variations of $t_D$, known as the Transit Duration Variations (TDVs), which 
can also be attributed to the gravitational perturbations.
 
The dynamical analysis of TTVs can yield a comprehensive characterization 
of orbits and also the planetary masses, or at least provide certain limits 
(Miralda-Escud\'{e} 2002, Agol et al. 2005, Holman \& Murray 2005). The TTVs 
have been primarily used to confirm and characterize the systems of 
multi-transiting planets (Holman et al. 2010, Lissauer et al. 2011, 
Steffen et al. 2012, Xie 2014), but they can also be used to detect and 
characterize non-transiting planets (Ballard et al. 2011, 
Nesvorn\'{y} et al. 2012, 2013, 2014, Dawson et al. 2014, Mancini et al. 2016). 
In fact, if the radial velocity measurements are not available, the 
observation and analysis of TTVs provide the only way currently available 
to determine masses. When the mass determination is combined with the 
planetary radius estimate from the transit light curve, this results in a 
planetary density estimate, and hence allows scientists to develop internal 
structure models (e.g. Guillot \& Gautier 2015).

In this work, we focus on the Kepler-46 system (KOI-872). The first analysis 
of the light curve of Kepler-46 showed the presence of a candidate planet, 
currently known as Kepler-46b, with a 33.6 day period and radius consistent 
with a Saturn-class planet (Borucki et al. 2011). Nesvorn\'{y} et al. (2012) 
measured and analyzed the TTVs and TDVs of 15 transits from \textit{Kepler} quarters 1-6. 
They found that the TTVs can be best explained if Kepler-46b is a member of a 
two-planet system with the planetary companion, Kepler-46c, having a 
non-transiting orbit on outside of Kepler-46b. The TTVs analysis provided two 
possible solutions for the orbital parameters of the two planets, but one of 
these solutions was discarded by Nesvorn\'{y} et al. because it produced 
significant TDVs, which were not observed. The preferred solution indicates 
a nearly coplanar ($I_{\mathrm{mut}}\simeq 1^{\circ}$) and nearly circular 
($e\simeq 0.01$) orbits, and $M_c=0.376^{+0.021}_{-0.019}$~$M_{\mathrm{J}}$. The mass of 
Kepler-46b was not constrained from TTVs/TDVs, but an upper limit 
of $M_b<6$~$M_{\mathrm{J}}$ was determined from the stability analysis.

Here, we reanalyzed the Kepler-46 system using 35 transits from \textit{Kepler} quarters. 
The transit analysis and dynamical fits were executed using \texttt{MultiNest} 
(Feroz et al. 2009, 2013), which is an efficient tool based on the Bayesian 
inference method. We were able to better determine the orbital parameters and 
masses of the two planets. The mass of Kepler-46b was determined here from TTVs 
for the first time. The paper is organized as follows: in Section 
\ref{cap:lightcurve}, we describe the light curve analysis to obtain the 
mid-transit times and transit parameters of Kepler-46b. In Section 
\ref{cap:method}, we discuss the dynamical analysis. 
The results are reported in Section \ref{cap:results}. The last
section is devoted to the conclusions.

\section{Light curve analysis and transit times} \label{cap:lightcurve}

\begin{table}[!t]
	\caption{The mid-transit times $t_C$ obtained for Kepler-46b from the analysis 
	of \textit{Kepler} Quarters 1-16(Model $\mathcal{M}_I$).}
	\label{tab:ttvobs}
    \centering
             \begin{tabular}{cl}
             \hline         
             \hline         
              Cycle & \hspace {0.8cm}$t_{C}$ (BJD)  \\ [0.5ex] 
             \hline         
              1  &  $ 55019.6993  \pm  0.0012$ \\
              2  &  $ 55053.2959  \pm  0.0010$ \\
              3  &  $ 55086.8640  \pm  0.0011$ \\
              4  &  $ 55120.4409  \pm  0.0018$ \\ 
              6  &  $ 55187.6880  \pm  0.0015$ \\ 
              7  &  $ 55221.3360  \pm  0.0016$ \\ 
              8  &  $ 55254.9066  \pm  0.0011$ \\ 
              9  &  $ 55288.4664  \pm  0.0012$ \\ 
              11 &  $ 55355.6667  \pm  0.0010$ \\ 
              12 &  $ 55389.3438  \pm  0.0011$ \\ 
	          13 &  $ 55422.9309  \pm  0.0017$ \\ 
              14 &  $ 55456.4875  \pm  0.0011$ \\ 
              15 &  $ 55490.0601  \pm  0.0009$ \\ 
              18 &  $ 55590.9283  \pm  0.0008$ \\ 
              19 &  $ 55624.5138  \pm  0.0008$ \\ 
              20 &  $ 55658.0838  \pm  0.0011$ \\ 
              21 &  $ 55691.6593  \pm  0.0008$ \\ 
              22 &  $ 55725.3035  \pm  0.0009$ \\ 
              23 &  $ 55758.9147  \pm  0.0009$ \\ 
              24 &  $ 55792.5562  \pm  0.0010$ \\
              25 &  $ 55826.1262  \pm  0.0011$ \\ 
              26 &  $ 55859.6845  \pm  0.0017$ \\ 
              27 &  $ 55893.2891  \pm  0.0011$ \\ 
              29 &  $ 55960.5704  \pm  0.0010$ \\ 
              31 &  $ 56027.7090  \pm  0.0010$ \\ 
              32 &  $ 56061.2821  \pm  0.0010$ \\ 
              33 &  $ 56094.8799  \pm  0.0010$ \\  
              35 &  $ 56162.1511  \pm  0.0011$ \\  
              36 &  $ 56195.7346  \pm  0.0014$ \\ 
              37 &  $ 56229.3033  \pm  0.0012$ \\ 
              38 &  $ 56262.8845  \pm  0.0013$ \\ 
              39 &  $ 56296.5319  \pm  0.0012$ \\ 
              40 &  $ 56330.1453  \pm  0.0010$ \\ 
              41 &  $ 56363.7757  \pm  0.0010$ \\ 
              42 &  $ 56397.3426  \pm  0.0010$ \\    
             \hline
             \end{tabular}                
\end{table}

Before a dynamical analysis can be conducted, the transit times and durations
of each event need to be inferred. This, in turn, requires us to first detrend
the \textit{Kepler} photometry and then fit the transits with an appropriate
model. For this task, we use a similar approach to that described in Teachey et
al. (2017), which is essentially a modified version of the approach described
in Kipping et al. (2013).

The broad overview is that we remove long-term trends from the \textit{Kepler} 
simple aperture photometry with the \texttt{CoFiAM} algorithm described in Kipping et al.
(2013). Essentially, \texttt{CoFiAM} is a cosine filter designed to not disturb the
transit of interest but remove present longer-period trends. A pre-requisite
for using \texttt{CoFiAM} is that the time and duration of each transit are already
approximately known. Since this system is known to exhibit strong variations in
both, this poses a catch-22 problem for us.

Following Teachey et al. (2017), we remedy the problem by conducting an initial
detrending using approximate estimates for the times and a fixed duration taken
from Nesvorn\'{y} et al. (2014). This data are then fitted (as we will describe
later) to provide revised estimates for the times and durations. These revised
times and durations are then used as inputs for a second attempt at detrending
the original \textit{Kepler} data using \texttt{CoFiAM} again. This iteration process
allows for self-consistent inference of the basic transit parameters.

Initially, we take into account the 40 transit events available 
for Kepler-46b (e.g. Holczer et al. 2016). However, one of such transit 
(corresponding to cycle 5) is found to have inadequately detrended data, which is 
identified by visual inspection of the light curve. Four other events 
(corresponding to cycles 0, 10, 16, and 28) are found to have insufficient 
temporal coverage\footnote{In particular, the transit of cycle 16 is marked with 
an outlier code $>32$ in Holczer et al. (2016).}.
There are several reasons that this might happen. At first, safe modes 
and data downlinks cause genuine gaps in the time series. At the next level,
any \textit{Kepler} time stamp with an error flag not set to 0 is removed by
\texttt{CoFiAM}. Then, \texttt{CoFiAM} puts a smoothed moving median through the 
data and looks for 3$\sigma$ outliers caused by sharp flux changes or some 
other weird behavior. Stitching problems among the quarters 
can also cause apparent sharp flux changes, and if we do not have enough usable 
data around the transit, then the event will also have data chopped this way. 
At last, the detrending of an event can fail giving a poor detrending, 
so we perform a second round of outlier cleaning that which may eventually lead to most of 
the data for that event being removed.
In the end, we keep only 35 transits for which visual inspection of the light curve
indicates that we would be able to get reliable time estimates (Table \ref{tab:ttvobs}).

The actual fitting process, which is ultimately repeated twice, is conducted
using \texttt{MultiNest} coupled to a Mandel \& Agol (2002) light curve model. We assume
freely fitted quadratic limb-darkening coefficients, using the prescription of
Kipping (2013), and allow each transit to have a unique mid-transit time but a
common set of basic shape parameters (i.e. a common duration). While 20 of the
transits were short cadence, 15 were long cadence requiring re-sampling using
the technique of Kipping (2010), for which we used $N_{\mathrm{resam}}=30$.

Since each transit requires a unique free parameter, this leads to a very large
number of free parameters in the final fit (35 just for the times alone). For
computational expedience, we split the light curve into 4 segments of 10
transits each, with 5 for the last segment. Each segment is independent of the
others, but assumes internally consistent shape parameters. Rather than freely
fitting each event, this allows for the transit template to be well constrained
such that the mid-transit time precision is improved. We refer to the above model 
of segments of transits as model $\mathcal{M}_S$.

As we did in Nesvorn\'{y} et al. (2014), we also try fitting each transit
completely independently, but fixing the limb-darkening coefficients to the
same as those used in Nesvorn\'{y} et al. (2014). This allows for TDVs to
be inferred as well as TTVs, but generally increases the credible interval of
the TTV posteriors due to the much weaker information about the transit shape
available. In what follows, we refer to this model as $\mathcal{M}_I$.


\section{Dynamical analysis of transit times} \label{cap:method}

Following Nesvorn\'{y} et al. (2012), we proceed by searching for a dynamical 
model that can explain the measured transit times. At variance with Nesvorn\'{y} et al.,
we do not fit the TTVs but we directly fit the mid-transit 
times, which is expected to be a more accurate procedure. The results 
presented in the following are based on the mid-transit times from 
model $\mathcal{M}_I$ (Table \ref{tab:ttvobs}). We have performed similar computations using model 
$\mathcal{M}_S$ and verified that the results are indistinguishable
within their 1-$\sigma$ uncertainties.
 
\begin{table}[!t]
	\caption{Prior distributions of 12 model parameters that we used to 
	obtain the dynamical fit. $\mathrm{U}[x,y]$ means a uniform distribution between 
	$x$ and $y$. The subindex $_P$ refers to any of the two planets.}
	\label{tab:priors}
    \centering
             \begin{tabular}{ll}
             \hline         
             \hline         
             Parameter      & Prior values  \\ [0.5ex] 
             \hline         
             \vspace{0.1cm}
             $M_P/M_\star$ & $\mathrm{U}[0,\,0.005]$  \\
             \vspace{0.1cm}
             $P_b$ (days) & $\mathrm{U}[33.5,\,33.7]$  \\
             \vspace{0.1cm}
             $P_c$ (days) & $\mathrm{U}[40,\,84]$  \\
             \vspace{0.1cm}
             $e_P$ & $\mathrm{U}[0,\,0.4]$ \\                
             \vspace{0.1cm}
             $b_c$ & $\mathrm{U}[1,\,20]$  \\
             \vspace{0.1cm}                   
             ${\varpi}_P$~($^\circ$) & $\mathrm{U}[0,\,360]$ \\
             \vspace{0.1cm}    
             $\lambda_c$~($^\circ$) & $\mathrm{U}[0,\,360]$ \\
             \vspace{0.1cm} 
             $\Omega_c-\Omega_b$~($^\circ$) & $\mathrm{U}[0,\,360]$   \\
             \vspace{0.1cm} 
             $\delta t$~(d) & $\mathrm{U}[-0.1,\,0.1]$\\ 
             \hline
             \end{tabular}                
\end{table}

To perform the dynamical fit, we assume a model
of two planet and we use a modified version of the
efficient symplectic integrator \texttt{SWIFT} (Levison \& Duncan 1994), 
adapted to record the mid-transit times of any transiting planet 
(Nesvorn\'{y} et al. 2013, Deck et al. 2014). \texttt{MultiNest} is then used to 
perform the model selection and to estimate the best-fit values of the dynamical 
parameters with their errors.

\texttt{MultiNest} applies the Bayes rule (Appendix \ref{append}) to determine the
values of the 12 parameters 
of the model that are necessary to reproduce the observed mid-transit times. 
The method provides the posterior distributions of these 12 parameters,
namely, the planet-over-star mass ratios (${M_b}/{M_\star}$, ${M_c}/{M_\star}$), 
the orbital periods ($P_b$, $P_c$), the eccentricities ($e_b$, $e_c$), the 
longitudes of periastron ($\varpi_b$, $\varpi_c$), the mean longitude and 
impact parameter of the non-transiting planet ($\lambda_c$, $b_c$), the 
difference of the nodal longitudes of the two planets ($\Omega_c-\Omega_b$), 
and the difference between the mid-transit time reference epoch 
($\tau=55053.2826$~BJD) and the mid-transit time of the nearest transit 
($\delta t$). This latter parameter gives us the information about 
the mean longitude of the transiting body at the reference epoch 
($\lambda_b=\lambda_{b,0}-2\pi\,\delta t/P_b$, where $\lambda_{b,0}$ is
the mean longitude at mid-transit time).
The initial value of the impact parameter of Kepler-46b ($b_b$) is fixed at 
the value determined from the transit fit. We use the transit reference 
system from Nesvorn\'y et al. (2012), 
where the reference plane is the plane defined by $b=0$, 
the origin of longitudes is at the line of sight, and
the nodal longitude of Kepler-46b ($\Omega_b$) is set to $270^{\circ}$. 
The stellar parameters are also adopted from Nesvorn\'y et al. (2012). 

The priors distributions of the 12 model 
parameters are given in Table \ref{tab:priors}. 
The distributions were chosen as uninformative with uniform priors.
The interval limits of these priors are based on results 
from Nesvorn\'y et al. (2012) that guarantee that the system is constituted
of a transiting and a non-transiting planet and is dynamically stable. 
The orbital period of the transiting body is well known from the transit fit,
so we only consider a very small range of priors around the known value. 
For the period of the non-transiting planet, we consider an interval of priors
that includes the two solutions found by Nesvorn\'y et al. (2012). 
The upper limit in eccentricities is set to only 0.4, since higher values 
cause the code to become too slow and do not provide additional solutions.
The upper limit of the impact parameter of the non-transiting planet 
corresponds to an inclination $i_P\sim 70^{\circ}$.

The calculation of the integrals involved in Eq.~(\ref{eq:evidence}) 
requires the use of a numerical method. 
\texttt{MultiNest} uses a multi-modal nested sampling technique to efficiently compute 
the evidence integral, and also provides the posterior distributions 
of the parameters. In our case, the likelihood function (see
Appendix \ref{append}) is defined as
\begin{equation}
 L(d|\theta, M) = \prod_{j=1}^{N} \frac{1}{\sqrt{2\pi\sigma_j^2}} 
    \exp\left(-\sum_{i=1}^{N} 
	\frac{\left(t_{C,i}^{\mathrm{obs}}-t_{C,i}^{\mathrm{cal}}\right)^2}{2\sigma_i^2}
	\right),
	\label{eq:likelihood}
\end{equation}
where $N=35$ is the number of transits; $t_{C,i}^{\mathrm{obs}}$ and $t_{C,i}^{\mathrm{cal}}$ 
are the observed 
(Table \ref{tab:ttvobs}) and calculated mid-transit times, respectively; 
and $\sigma_i$ is the uncertainty of $t_{C,i}^{\mathrm{obs}}$. 
According to Eq.~(\ref{eq:bayes}), the best-fit parameters are obtained by 
maximizing this likelihood function.

\section{Results}\label{cap:results}

Our analysis provides two best-fit solutions that correspond to the 
two solutions reported in Nesvorn\'{y} et al. (2012). The first solution 
(S1) was obtained for a uniform distribution of the period prior in 
the interval $P_c=[40,70]$~days. This solution has a maximum likelihood of 
$\ln L_{\mathrm{S1}}=183.42$, 
a reduced $\chi^2_{\mathrm{S1}}=1.96$ (for $35-12=23$ degrees of freedom), and 
a global evidence of $\ln Z_{\mathrm{S1}}=125.23$. The reduced $\chi^2$
indicates that in principle this is a good fit, but the transit time errors may 
be underestimated by a factor of $\sim 0.8$.

\begin{figure}[!p]
    \centering
	\includegraphics[width=0.55\columnwidth]{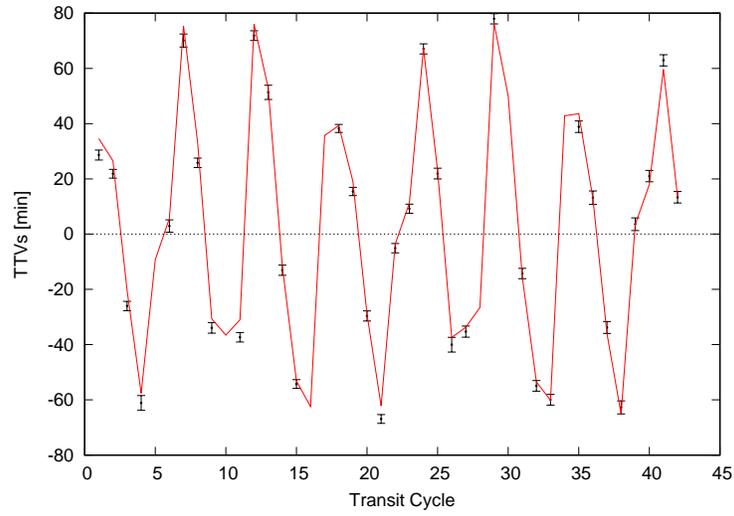}
	\caption{Transit Timing Variations (black dots) computed as the difference 
	between the observed mid-transit times and the linear ephemeris (obtained 
	by fitting a straight line to the mid-transit times series). 
	The red line corresponds to the best dynamical fit corresponding to the solution S1.}
    \label{fig:O-C}
\end{figure}

Table \ref{tab:param} reports the parameter values corresponding to S1. 
They were determined from the weighted posteriors calculated by \texttt{MultiNest}. 
The associated uncertainties are defined by the standard error at the 68.2\% 
confidence level. The orbital elements provided by this solution are 
astrocentric osculating elements at epoch BJD~55053.2826.
The corresponding fit to the TTV signal is shown in 
Fig. \ref{fig:O-C}.

\begin{table}[!t]
	\caption{The best-fit parameters and their errors for the solution S1. 
	The first block reports the transit parameters of Kepler-46b obtained from 
	the light-curve analysis. The second block reports the orbital parameters and
	masses obtained from the dynamical fit. 
	Orbital elements are astrocentric at osculating epoch BJD 55053.2826,
	the reference plane is the plane at which the impact parameter is 0,
	and the origin of longitudes is at the line of sight.
	The third block reports the secondary parameters. $i_P$ is the inclination 
	of the planet with respect to the sky plane 
	during transit ($\cos i_P\simeq b_P R_\star /a_P$); 
	$I_p$ is the inclination of the planetary orbit with respect to  
	the reference plane; 
	$I_{\mathrm{mut}}$ is the mutual inclination relative to the orbital
	plane of Kepler-46b. The last block lists the stellar parameters 
	compiled from Nesvorn{\'{y}} et al. (2012).}
	
	\label{tab:param}
    \centering  
             \begin{tabular}{lll}
             \hline         
             \hline         
             & \bf{Kepler-46b}  &  \bf{Kepler-46c} \\ [0.5ex] 
             \hline         
             Transit fit            &  &    \\ [0.5ex] 
                \vspace{0.1cm}
                $R_P/R_\star$ & $0.0887^{+0.0010}_{-0.0012}$ & \\
                \vspace{0.1cm}
                $b_P$ & $0.757^{+0.022}_{-0.027}$ &  \\
             \hline         
             Dynamical fit            &  &     \\ [0.5ex] 
                \vspace{0.1cm}
                $M_p/M_\star$~($\times 10^{-4}$) & $9.372^{+3.941}_{-3.618}$ & $3.835^{+0.057}_{-0.055}$ \\
                \vspace{0.1cm}
                $P_P$~(days) & $33.648^{+0.004}_{-0.005}$ & $57.325^{+0.116}_{-0.098}$ \\
                \vspace{0.1cm}
                $e_P$ & $0.0321^{+0.0069}_{-0.0078}$ & $0.0354^{+0.0057}_{-0.0059}$ \\                
                \vspace{0.1cm}
                $b_P$ & $0.757^{+0.022}_{-0.027}$ & $1.483^{+0.418}_{-0.322}$ \\
                \vspace{0.1cm}                   
                $\varpi_P$~($^\circ$) & $264.2^{+8.2}_{-8.9}$ &  $294.16^{+8.70}_{-6.42}$\\
                \vspace{0.1cm}    
                $\lambda_P$~($^\circ$) & -- & $338.0^{+0.3}_{-0.3}$ \\
                \vspace{0.1cm} 
                $\Omega_P$~($^\circ$) & $270$  & $261.4^{+22.7}_{-24.3}$ \\
                \vspace{0.1cm} 
                $\delta t$~(d) & $0.0130^{+0.0006}_{-0.0006} $  & -- \\    
             \hline         
             Secondary parameters    &  &     \\ [0.5ex] 
                \vspace{0.1cm}                
                $M_P$~($M_{\mathrm{J}}$) & $ 0.885^{+0.374}_{-0.343} $ & $0.362^{+0.016}_{-0.016}$  \\
                \vspace{0.1cm}                               
                $a_P$~(au) & $0.1971^{+0.0001}_{-0.0001}$ & $0.2811^{+0.0003}_{-0.0003}$ \\     
                \vspace{0.1cm} 
                $i_P$~($^\circ$) & $89.04^{+0.14}_{-0.14}$ & $88.66^{+0.26}_{-0.27}$ \\   
                \vspace{0.1cm}                                 
                $I_P$~($^\circ$) & $0.957^{+0.028}_{-0.034}$ & $1.35^{+0.38}_{-0.29}$ \\   
                \vspace{0.1cm}                                 
                $I_{\mathrm{mut}}$~($^\circ$) & -- & $0.43^{+0.40}_{-0.26}$ \\   
                \vspace{0.1cm}                                 
                $R_P$~($R_{\mathrm{J}}$) & $0.810^{+0.035}_{-0.036}$ & -- \\
                \vspace{0.1cm}
                $\rho$~(g\,cm$^{-3}$) &  $2.069^{+0.913}_{-1.136}$ & -- \\
             \hline         
             & \multicolumn{2}{c}{\bf{Kepler-46}}  \\ [0.5ex] 
             \hline         
             Stellar parameters & &  \\  [0.5ex]
               \vspace{0.1cm}
               $M_\star$~($M_\odot$) & \multicolumn{2}{c}{$0.902^{+0.040}_{-0.038}$}  \\
               \vspace{0.1cm}
               $R_\star$~($R_\odot$) & \multicolumn{2}{c}{$0.938^{+0.038}_{-0.039}$}  \\
               \vspace{0.1cm}
               $\rho_\star$~(g\,cm$^{-3}$) & \multicolumn{2}{c}{$1.54^{+0.22}_{-0.17}$}  \\
               \vspace{0.1cm}
               $\log g_\star$\tablenotemark{a} & \multicolumn{2}{c}{$4.447^{+0.040}_{-0.035}$}  \\
               \vspace{0.1cm}
               $T_{\mathrm{eff}}$~(K) & \multicolumn{2}{c}{$5155^{+150}_{-150}$}  \\
               \vspace{0.1cm}
               $L_\star$~($L_\odot$) & \multicolumn{2}{c}{$0.556^{+0.078}_{-0.070}$}  \\
               \vspace{0.1cm}
               $M_V$  & \multicolumn{2}{c}{$5.60^{+0.17}_{-0.17}$}  \\
               \vspace{0.1cm}
               Age~(Gyr) & \multicolumn{2}{c}{$9.7^{+3.7}_{-3.5}$}  \\
               \vspace{0.1cm}
               Distance~(pc) & \multicolumn{2}{c}{$855^{+68}_{-65}$}  \\
               \vspace{0.1cm}
               [M/H]  & \multicolumn{2}{c}{$0.41^{+0.10}_{-0.10}$}  \\
               \hline
               \end{tabular} 
			   \tablenotetext{\textrm{a}}{$g_\star$ is given in c.g.s units.}               
\end{table}

\begin{table*}[!t]
	\caption{Similar to Table \protect\ref{tab:param} but for the two modes of 
	solution S2.}
	\label{tab:param2}         
    \centering
            \begin{tabular}{lllll}
            \hline         
            \hline         
            & \multicolumn{2}{c}{Mode 1}  & \multicolumn{2}{c}{Mode 2} \\ [0.5ex]
            & \bf{Kepler-46b}  &  \bf{Kepler-46c} & \bf{Kepler-46b} & \bf{Kepler-46c}  \\ [0.5ex] 
            \hline
            Dynamical fit & & & & \\    [0.5ex]        
            \vspace{0.1cm}
            $M_p/M_\star$~($\times 10^{-4}$) & $8.223^{+20.522}_{-6.530}$ & $20.151^{+0.930}_{-0.930}$ & $7.499^{+19.497}_{-5.875}$ & $20.148^{+0.890}_{-0.957}$ \\                  
            \vspace{0.1cm}
            $P_P$~(d) & $33.609^{+0.002}_{-0.003}$ & $81.504^{+0.165}_{-0.501}$ & $33.610^{+0.002}_{-0.003}$ & $81.520^{+0.152}_{-0.479}$ \\
            \vspace{0.1cm}
            $e_P$ & $0.0063^{+0.0047}_{-0.0040}$ & $0.0239^{+0.0031}_{-0.0033}$ & $0.0067^{+0.0051}_{-0.0042}$& $0.0240^{+0.0032}_{-0.0034}$ \\                
            \vspace{0.1cm}
            $b_P$ & $0.757$ & $4.962^{+1.983}_{-1.153}$ & $0.757$ & $7.453^{+1.854}_{-1.175}$ \\   
            \vspace{0.1cm}
            $\varpi_P$~($^\circ$) & $225.9^{+42.3}_{-108.2}$ &  $124.1^{+7.8}_{-12.8}$& $229.3^{+38.6}_{-102.3}$ & $124.0^{+7.9}_{-13.2}$  \\
            \vspace{0.1cm}
            $\lambda_P$~($^\circ$) & -- & $197.2^{+2.6}_{-2.5}$ & -- & $196.7^{+2.7}_{-2.6}$ \\               
            \vspace{0.1cm}
            $\Omega_P$~($^\circ$) & $270$  & $19.4^{+8.7}_{-4.5}$  & $270$ & $207.7^{+7.5}_{-4.1}$ \\ 
            \hline         
            Secondary parameters & & & & \\   [0.5ex]        
            \vspace{0.1cm}
            $M_P$~($M_{\mathrm{J}}$) &  $0.777^{+1.939}_{-0.617}$  & $1.904^{+0.121}_{-0.119}$ & $0.708^{+1.842}_{-0.562}$ & $1.886$  \\
            \vspace{0.1cm}
            $a_P$~(au) & $0.1969^{+0.0001}_{-0.0001}$ & $0.3556^{+0.0006}_{-0.0006}$  & $0.1969^{+0.0001}_{-0.0001}$ & $0.3557^{+0.0006}_{-0.0006}$ \\                    
            \vspace{0.1cm}
            $i_P$~($^\circ$) & $89.04^{+0.14}_{-0.14}$ & $86.51^{+1.85}_{-1.90}$ & $89.04^{+0.14}_{-0.14}$ & $84.75^{+4.31}_{-4.20}$ \\                                    
            \vspace{0.1cm}
            $\rho$~(g\,cm$^{-3}$) & $1.815^{+4.537}_{-2.042}$ & -- & $1.655^{+4.310}_{-1.858}$ & -- \\
            \hline
            \end{tabular}
\end{table*}

\begin{figure*}[!p]
    \centering
    \includegraphics[width=\textwidth]{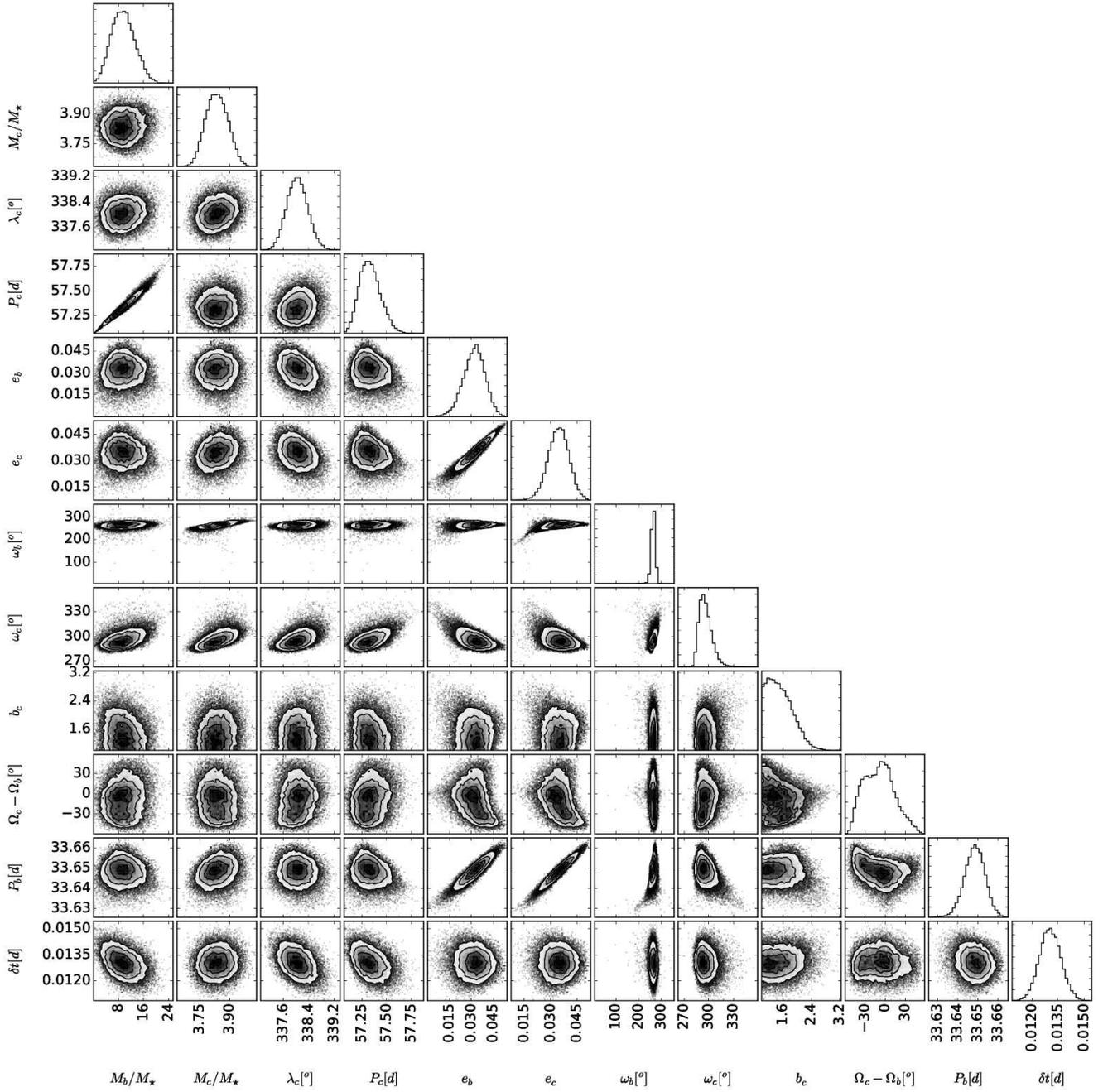}
	\caption{The equally weighted posterior distributions of our 12 model 
	parameters (diagonal), and the corresponding correlations between 
	parameters. The mass ratios are given in units of} $10^{-4}$.
    \label{fig:times2}
\end{figure*}

The posterior distributions of parameters, and the correlations between 
pairs of parameters, are shown in Fig. \ref{fig:times2}. According to solution S1, 
Kepler-46b is a Jupiter class planet ($M_b\simeq 0.9$~$M_{\mathrm{J}}$), with a density 
similar or slightly higher than that of Jupiter. The companion is a Saturn 
class planet ($M_c\simeq 0.36$~$M_{\mathrm{J}}$)\footnote{%
This mass value is the same as determined by Nesvorn{\'{y}} et al. (2012), 
and it is in good agreement with the analytic estimate of Deck \& Agol (2016).} 
moving on an outer orbit with 
$P_c=57.325$~d.

The period ratio ${P_c}/{P_b}=1.703$ indicates that the two planets are 
close to the 5:3 mean motion resonance, but their orbits are not resonant. 
This configuration may be part of the trend identified previously where compact 
planetary systems appear to avoid exact resonances (e.g., Veras \& Ford 2012). 
At least in some cases, this can be explained by tidal dissipation 
acting in the innermost planet (Lithwick \& Wu 2012, Batygin \& Morbidelli 2013). 
Kepler-46b, however, is probably too far from its host star for the tidal effects 
to be important. The orbital eccentricities of the two planets are similar, 
$e_b,e_c \simeq 0.03$, and the mutual inclination of the orbits is 
$I_{\mathrm{mut}}=0.43^{\circ}$, confirming the nearly circular and nearly coplanar 
nature of the planetary system. The small mutual inclination
is indicates that Kepler-46c may become a transiting planet in 
the near future (see Section \ref{longterm}).

The second solution (S2) is obtained for the period prior $P_c=[80,84]$~days. 
This solution has a global evidence of $\ln Z_{\mathrm{S2}} = 105.97$ and reveals the 
existence of two modes differing mainly in the mutual inclination between orbits. 
Specifically, Mode 1 of S2 corresponds to an impact parameter 
for Kepler-46c of $b_c\simeq 5$ with $0^{\circ}\leq\Omega_c \leq180^{\circ}$,
while Mode 2 corresponds to $b_c\simeq 7$ with $180^{\circ}\leq\Omega_c \leq360^{\circ}$. 
Mode 1 gives a mutual inclination of $I_{\mathrm{mut}}\simeq 170^{\circ}$ 
implying that the orbit of Kepler-46c is retrograde, while 
Mode 2 gives $I_{\mathrm{mut}}\simeq 10^{\circ}$ meaning that both planets
are in prograde orbits. The maximum likelihood 
and reduced $\chi^2$ values of the two modes are: $\ln L_{\mathrm{S2}}=157.16$,
$\chi^2_{\mathrm{S2}}=4.25$ and $\ln L_{\mathrm{S2}}=157.56$, $\chi^2_{\mathrm{S2}}=4.21$, respectively.
Table \ref{tab:param2} summarizes some characteristics of the solution S2.

For both modes, the mass of Kepler-46c would be larger ($M_c\simeq 1.9$~$M_{\mathrm{J}}$) 
than that obtained for the solution S1. The orbital eccentricities of both 
modes are smaller than for the S1 solution. The orbital period ratio suggested 
by both modes is ${P_c}/{P_b} \simeq 2.42$, which places the two planets very 
close to (but not inside of) the 5:2 mean motion resonance.

Since a $\Delta\chi^2\sim 2.25$ between solutions S1 and S2 is not statistically significant enough
to penalize any of the solutions, we apply Eq.~(\ref{eq:posteriorodss})
to compare the global evidences of S1 and S2.
The Bayes factor becomes $\Delta \ln Z=19.26$, suggesting that 
solution S1 is preferred over S2 with a confidence level of 5.7$\sigma$. This argument can be 
used to rule out solution S2. It is worth stressing that this conclusion is based on the 
analysis of the TTVs only, while Nesvorn\'{y} et al. (2012) had to resort 
to using the TDV constraints to arrive at the same conclusion. 

In order to test how sensitive is the Bayes factor is to the choice of prior distributions,
we redo the dynamical fits restricting the intervals of the $e_b,e_c$ priors to
$\mathrm{U}[0,\,0.2]$ and the interval of the $P_c$ priors to $\mathrm{U}[55,\,84]$. 
The resulting values of the evidence $\ln Z$ 
of each solution are higher in this case, as expected from 
Eq.~(\ref{eq:evidence}), but the Bayes factor is almost the same: 
$\Delta \ln Z=20.81$. This indicates that S1 is still preferred over S2
with a confidence level of 6$\sigma$. Then, we can claim that model selection relying on
the Bayes factor is not sensitive to the choice of the priors distributions.

Finally, it is worth noting that if we perform the dynamical fits
using more than 35 transits, the solutions do not change significantly. 
At least three of the five discarded transits (corresponding to cycles 0, 5, and 10;
see Sect. \ref{cap:lightcurve}) allow us to estimate mid-transit times, 
even with large errors.
The dynamical fits using these $35+3$ transits provide again the two solutions S1 
and S2, with S2 showing two modes, and the Bayes factor favoring S1. 
All the parameters of these fits are indistinguishable within 
1-$\sigma$ errors with respect to the parameters reported in Tables 
\ref{tab:param} and \ref{tab:param2}. 
We conclude that our two-planet model solution S1 is quite robust.

\subsection{Long-term stability}\label{longterm}

Nesvorn\'{y} et al. (2012) demonstrated that their S1 solution is dynamically 
stable over 1~Gyr. Since here we obtained slightly larger values of the orbital 
eccentricities, slightly smaller values of the mutual inclination, and were able 
to constrain the mass for Kepler-46b, we find it useful to reanalyze the long-term 
stability of the system. 

\begin{figure}[!p]
    \centering
	\includegraphics[width=0.55\columnwidth]{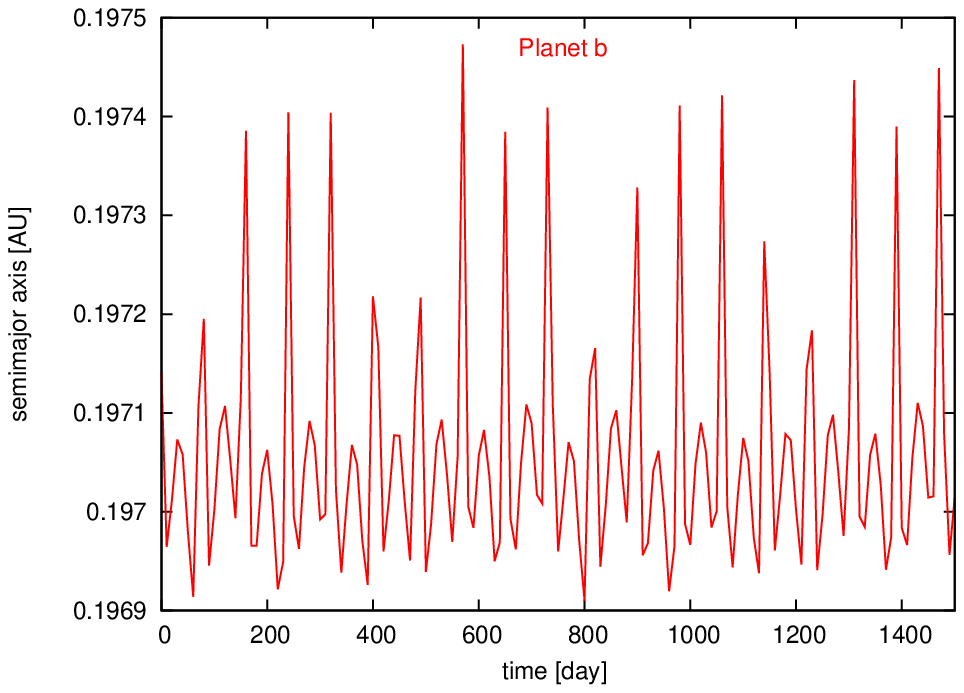}
	\includegraphics[width=0.55\columnwidth]{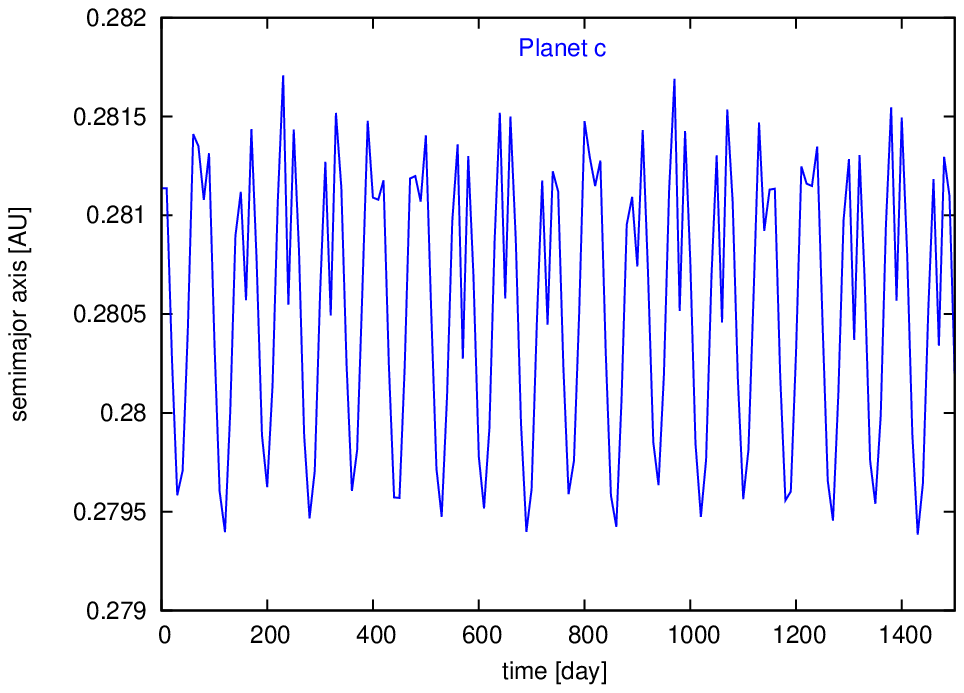}
	\caption{Evolution of the semi-major axes of Kepler-46b (top panel) 
	and Kepler-46c (bottom panel)}
    \label{fig:a_kep46}
\end{figure}

\begin{figure}[!p]
    \centering
	\includegraphics[width=0.55\columnwidth]{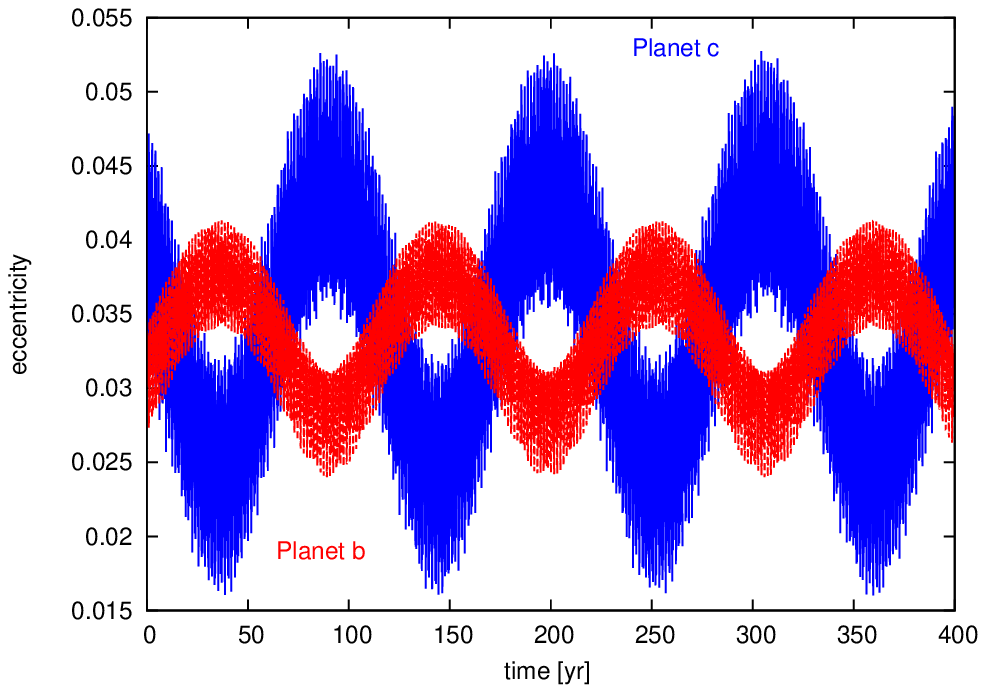}
	\includegraphics[width=0.55\columnwidth]{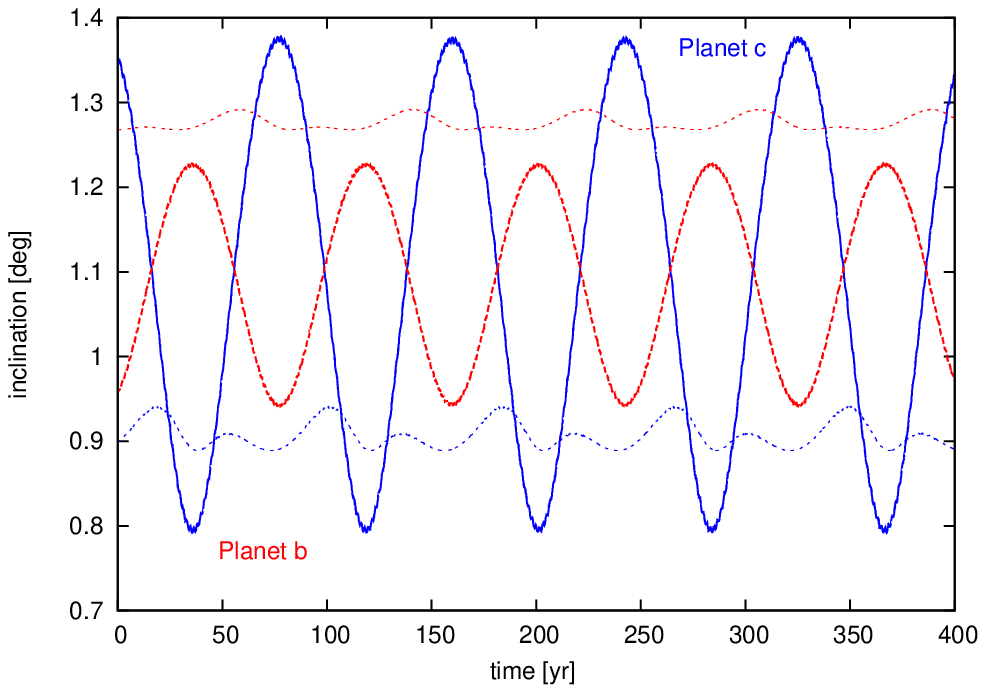}
	\caption{Evolution of the eccentricities (top panel) and inclinations 
	(bottom panel) of Kepler-46b (red full lines) and Kepler-46c (blue full lines)
	The inclinations are relative to the transit reference plane. 
	The dotted lines in the bottom panel represent the inclination limits 
     below which the planets are expected to show transits (i.e. $b_P<1$).}
    \label{fig:e-inc_kep46}
\end{figure}

We used the orbits and masses corresponding to the solution S1 (see Table \ref{tab:param}) 
and numerically integrated the orbits over 100~Myr using the \texttt{SWIFT} code and a one-day 
time step. The orbital evolution of the planets in the first 1500 days of this integration 
is shown in Figs. \ref{fig:a_kep46} and \ref{fig:e-inc_kep46}. 
The semi-major axes show short-term oscillations with amplitudes never larger 
than $0.7\%$. Due to the conservation of the total angular momentum, the 
eccentricities show anti-correlated oscillations with a period of $\simeq 110$~years. 
We note that, for both planets, these oscillations show amplitudes that are of the same 
order of the eccentricity uncertainties reported in Table \ref{tab:param}.
This implies that the eccentricities will be indistinguishable at any epoch, within
their 1$\sigma$ uncertainties.
The inclinations $I_P$ relative to the transit reference plane 
(see Table \ref{tab:param}) also show anti-correlated 
oscillations with a period of $\simeq 83$~yr resulting from the precession of 
the nodal longitudes. In this case, the oscillation amplitude of
Kepler-46c is also of the same order of the uncertainty in inclination, but for
Kepler-46b the amplitude is about five times larger than the uncertainty.
The integration results indicate that the orbits show no 
signs of chaos, and the planetary system is stable at least over 100~Myr.

Our analysis also shows that Kepler-46b should always transit the star, as can 
be seen from the right panel of Fig. \ref{fig:e-inc_kep46}. Kepler-46c, on 
the other hand, is not currently transiting, but it may start to display 
transits in a few decades. Once it does, Kepler-46 will be a good target for 
light curve observations that may lead to the confirmation of Kepler-46c and the 
determination of its radius and density. 

\section{Conclusion}

We reanalyzed the Kepler-46 planetary system using a larger number of 
transits (35) than in Nesvorn\'{y} et al. (2012) and applying 
the Bayesian inference to perform a dynamical fit to the measured TTVs. 
We obtained two possible solutions, but the Bayesian evidence allows us 
to rule out one of them without the use of TDVs as in Nesvorn\'{y} 
et al. (2012). The availability of a larger number of transits allows us 
to determine the mass Kepler-46b, and constrain its density.

We confirm that the TTVs signal is well reproduced by a system of two 
planets on nearly circular and coplanar orbits, with periods of
$P_b \simeq 33.6$~days and $P_c \simeq 57.3$~days, respectively. This means 
that the two planets are close to, but not inside of, the 5:3 mean 
motion resonance. With the radius of Kepler-46b determined from the 
photometric light curve, $R_b \simeq 0.85$~$R_{\mathrm{J}}$, and the new 
mass determination $M_b \simeq 0.9$~$M_{\mathrm{J}}$, we can constrain its 
density to be $\sim 2$~g\,cm$^{-3}$. This indicates, consistently with 
its Jupiter-like mass, that Kepler-46b has a significant gas component. 
No density estimate can be inferred for the non-transiting companion 
Kepler-46c, but the estimated mass $M_c \simeq 0.36$~$M_{\mathrm{J}}$ indicates a 
Saturn-class planet. Interestingly, our new fit indicates that Kepler-46c 
should start transiting the host star in a few decades.

\appendix

\section{Bayesian inference}\label{append}

In order to find the best-fit parameters to the observed mid-transit times, 
we apply a powerful statistics tool that relies on the Bayes rule:
\begin{equation}
 P(\theta|d,M)=\frac{L(d|\theta,M)\pi(\theta,M)}{Z(d|M)}
 \label{eq:bayes}
\end{equation}
This rule gives the posterior distribution $P(\theta|d, M)$ of the parameters 
$\theta$, for the model $M$, and gives the data $d$, in terms of the 
likelihood distribution function $L(d|\theta, M)$ within a given set of 
prior distribution $\pi(\theta, M)$. The expression is normalized by 
the so-called Bayesian evidence $Z(d|M)$. We recall that in this formula,
the prior and the evidence represent probability distributions, while the 
likelihood is a function that generate the data $d$ given the parameter 
$\theta$. The prior is the probability of the parameters $\theta$ that 
is available before making any observation, in other words, it is our 
state of knowledge of the parameters of the model before considering any 
new observational data $d$. The evidence is the probability of the data $d$ 
given the model $M$, integrated over the whole space of parameter $\theta$ 
as defined by the prior distribution. This is also referred to as the 
marginal likelihood, and it is given by
\begin{equation}
 Z(d|M)=\int{ L(d|\theta,M)\pi(\theta,M)}d\theta
 \label{eq:evidence}
\end{equation}

For parameter estimation, the Bayesian evidence simply enters as a 
normalization constant that can be neglected, as in most cases it seeks 
to maximum posterior values. However, in the case of having competing models, 
Bayes rule penalizes them via model selection, and in this case, the Bayesian 
evidence is relevant. For example, consider two models $M_1$ and $M_2$; 
then, the ratio of the posterior probabilities, also known as 
\textit{posterior odds}, is given by
\begin{equation}
 \frac{P(M_1|d)}{P(M_2|d)}=\frac {Z(d|M_1)}{Z(d|M_2)}\frac{\pi(M_1)}{\pi(M_2)}
 \label{eq:posteriorodss}
\end{equation}
Therefore, the posterior odds of the two models is proportional to the 
ratio of their respective evidences, which is called the \textit{Bayes factor}. 
The proportionality becomes an equality when the ratio of the prior probabilities 
(or \textit{prior odds}) of the two models is 1.

\acknowledgments

This work has been supported by the Coordination for the Improvement 
of Higher Education Personnel (CAPES), the Brazilian Council of Research
(CNPq), NASA's Emerging Worlds program, and Brazil's
Science without Borders program.

\end{document}